\documentclass[epj,final]{svjour}
\usepackage{graphicx}
\begin{document}

%\baselineskip=1. \baselineskip
%\draft
\title{Nuclear matter with off-shell
propagation}
\baselineskip=1. \baselineskip

\author{     P. Bo\.{z}ek%\footnote{electronic
%address~
%piotr.bozek@ifj.edu.pl}
}
\institute{
Institute of Nuclear physics, PL-31-342 Cracow, Poland}

\date{\today}

%\begin{abstract}
\abstract{
Symmetric nuclear matter is studied within the conserving, self-consistent 
$T$-matrix approximation. This approach involves off-shell propagation of
nucleons in the ladder diagrams. The binding energy receives contributions
from the background part of the spectral function, away form the
quasiparticle peak. The Fermi energy at the saturation point fulfills the
Hugenholz-Van Hove relation. In comparison to the Brueckner-Hartree-Fock
approach, the
binding energy is reduced and the equation of state is harder.
%{ Keywords :}{
%Nuclear matter, superfluidity, spectral function}
%\end{abstract}
}
%\maketitle

%\vskip .3cm

%\keywords

%\vskip .3cm

%\keywords{ 
%{Nuclear matter,% saturation point, thermodynamic properties,
% Hugenholz-Van Hove theorem}

%\vskip .4cm

%\pacs
%{\bf  21.65+f, 24.10Cn, 26.60+c}

%\vskip .4cm

\maketitle

The calculation of nuclear matter
properties from the basic nucleon-nucleon interaction has been 
extensively studied using Brueckner
type resummation of ladder diagrams. 
This resummation allows to rewrite the ground-state energy of nuclear matter
using as  an effective interaction, the $G$ matrix, which
takes care of the short range repulsive core in the nucleon-nucleon 
interaction  \cite{jlm}.
Calculations  using realistic interactions
lead to results, which lie
 along a line (the Coester line)
shifted with respect to the phenomenological saturation point
($\rho_0\simeq.16{\rm fm}^{-3}$, $E/N\simeq-16$MeV).
The remaining discrepancy %has been the subject of much discussion
%\cite{bhf}. 
 can be attributed to relativistic effects and three-body 
forces contributions \cite{bhf}.

The results on the binding energy depend  on the single
particle energies used in the kernel of the Bethe-Goldstone
equation \cite{bg}. The so called standard choice uses a self-consistent
auxiliary
potential defined by the $G$-matrix below the Fermi energy and the free
dispersion relation above $k_F$. Another choice is to use the self-consistent 
potential also above the Fermi momentum which gives the so called continuous
choice for the single-particle energies in the Bethe-Goldstone equation.
In Brueckner-Hartree-Fock (BHF) calculations the hole line expansion, 
irrespective of the choice of the auxiliary potential, is believed to converge 
to values close to the BHF with the continuous choice for single-particle 
energies \cite{baldo}.
%The resulting dispersion relation for the continuous choice has no gap at the
%Fermi momentum.
% but one has to neglect the imaginary part of the
%self-consistent potential above $k_F$.

Recently self-consistent approaches based on the in medium $T$-matrix 
approximation for nuclear matter have been studied 
\cite{di1,ja1,ja2,gent,jc,ja3}. In this way a spectral function for nucleons in
nuclear matter  including two-particle correlations is obtained.
 The ladder diagrams 
involved in the calculation of the in medium $T$ matrix include also 
hole-hole propagation. 
 The $T$-matrix approximation takes into account some of the higher order hole 
line contributions as compared to the $G$-matrix approach.
It would be instructive to study the saturation 
properties of nuclear matter for the self-consistent $T$-matrix approximation
with realistic interactions.

The $T$-matrix approach is a $\Phi$-derivable approximation 
\cite{baym}.
The self-energy is constructed as a functional derivative of a set of 
two-particle irreducible diagrams. 
This assures the fulfillment of thermodynamical relations for the 
quantities obtained \cite{jc}.
The most famous such a relation is the equality of the Fermi energy 
and binding energy at the saturation point \cite{hvh}
\begin{equation}
\label{hvheq}
E_F=E/N \ .
\end{equation}
The realization of the above relation %was  a kind of a holly grail quest for
%nuclear matter practitioners, 
is very important
since it would give confidence to the
single particle properties obtained in the calculations. 
In Ref. \cite{jc} we studied the 
self-consistent $T$-matrix approximation with a simple interaction 
confirming to a very good accuracy the 
fulfillment of  thermodynamical relations by the numerical solutions.
In BHF calculations the Hugenholz-Van Hove relation is badly violated. 
This discrepancy can be reduced by invoking rearrangement terms for the
Fermi energy
 \cite{gmhvh,jong,molinari}. By construction,
the single-particle energies obtained in
 the $T$-matrix approximation   come out 
consistently with thermodynamical observables. Thus
 we expect that single-particle energies, scattering width or spectral 
functions directly obtained from the self-consistent $T$-matrix 
approximation are
meaningful \cite{ja3}.

%  However in violation of the Hugenholz-Van Hove theorem
%the resulting Fermi energy $E_F$
% at saturation point is usually very different from 
%the binding energy per particle $E/N$. It is a manifestation of a general 
%violation of thermodynamic consistency by the $G-$matrix approximation.
%The problem was discussed in the literature
% and improvements due
%to rearrangement terms were used but without removing the discrepancy 
%altogether.

%Calculations involving off-shell propagators in the $T-$matrix ladder
%have been recently performed \cite{dickhoff,ja,ja2,gent} both in the
% normal and in the superfluid phase. Below we shall restrict ourselves to 
%zero temperature normal nuclear matter.

For attractive interactions cold nuclear matter forms a superfluid.
%The superfluid transition at the critical temperature 
%is directly related to the appearance of a singularity in the $T$-matrix
%at zero total momentum, and the energy of the pair equal to twice the 
%Fermi energy \cite{thouless,ro1,ja1,ja2}. %The presence 
%of such a singularity is equivalent to the existence of a non-zero solution 
%of the gap equation. The use of dressed propagators in the $T$-matrix, or in 
%the gap equation, reduces considerably the critical temperature \cite{ja1}.
Calculations using dressed propagators 
in the superfluid phase show a strong reduction 
of the gap \cite{ja2,ja4,jc2}. %The smaller value of the superfluid 
%gap  can be understood as due to 
% a reduction of the effective interaction and a renormalization of the 
%effective energy gap \cite{ja3,cs}. 
We  expect that around the saturation point the superfluidity 
is very weak \cite{ja4}. This means that
 the correction from the superfluid 
correlation   energy to the binding energy is small.
We restrict ourselves to normal nuclear matter for all 
densities studied here. It allows us to compare with BHF calculations which are
performed exclusively in the normal phase of  nuclear matter.

The results here presented are obtained using a separable parameterization 
of the Paris potential \cite{hp} for $S$, $P$, $D$ and $F$ partial waves,
for symmetric nuclear matter.
We use rank $3$ and rank $4$ parameterization for the $^1S_0$ and $^3S_1-^3D_1$
partial waves. In the $^3P_0$ partial wave we use Mongan I interaction, in
order to avoid unphysical resonances far off-shell.
 In the numerical iteration the full spectral 
function  is discretized. For momenta close to the Fermi momentum the 
spectral function is separated into a background part and a quasiparticle peak
approximated by a delta function.
The numerical treatment of the energy integrations
 for the spectral functions is done using convolution algorithms \cite{ja3}.

The $T$-matrix approximation resumes ladder diagrams with dressed 
particle-particle and hole-hole propagators 
\begin{eqnarray}
\label{teq}
& & <{\bf p}|T({\bf P},\Omega)|{\bf p}^{'}> = V({\bf p},{\bf p}^{'})
\nonumber \\ & & + 
 \int\frac{d\omega_1}{2\pi}\int\frac{d\omega_2}{2\pi}
\int\frac{d^3q}{(2 \pi)^3} V({\bf p},{\bf q})% \nonumber \\ 
%& & 
\frac{\big(1-f(\omega_1)-f(\omega_2)\big)}
{\Omega-\omega_1-\omega_2+i\epsilon} \nonumber \\ & &
A(p_1,\omega_{1})A(p_2,\omega_{2})
 <{\bf q}|T({\bf P},\Omega)
|{\bf p}^{'}> 
\end{eqnarray}
where ${\bf p_{1,2}}={\bf P}/2\pm {\bf q}$~ and $f(\omega)$ is the Fermi 
distribution.
The imaginary part of the corresponding retarded self-energy can be obtained
closing a pair of external vertices in the $T$-matrix with a fermion 
propagator
\begin{eqnarray}
\label{imags}
& & 
{\rm Im}
\Sigma(p,\omega) =\int\frac{d\omega_1}{2 \pi}\int \frac{d^3k}{(2 \pi)^3}
A(k,\omega_1) \nonumber \\ 
& & <({\bf p}-{\bf k})/2|{\rm Im}T({\bf p}
+{\bf k},\omega+\omega_1)|({\bf p}-{\bf k})/2>_A \nonumber \\ & & 
 \Big( f(\omega_1)+b(\omega+\omega_1) \Big) 
\end{eqnarray}
where 
\begin{equation}
\label{spectralf}
A(p,\omega)=\frac{-2 {\rm Im}\Sigma(p,\omega)}{\left(\omega-p^2/2m 
-{\rm Re}\Sigma(p,\omega)\right)^2 +{\rm Im}\Sigma(p,\omega)^2}
\end{equation}
is the self-consistent spectral function of the nucleon and $b(\omega)$
is the Bose distribution.
The real part of the self-energy is related to ${\rm Im} \Sigma$
by a dispersion relation
\begin{equation}
\label{reals}
{\rm Re}\Sigma(p,\omega)= \Sigma_{HF}(p) + {\cal P} \int \frac{d\omega^{'}}
{\pi} \frac{{\rm Im}\Sigma(p,\omega^{'})}{\omega^{'}-\omega}
\end{equation}
with $\Sigma_{HF}(p)$ the Hartree-Fock self-energy.
Eqs. \ref{teq}, \ref{imags}, \ref{reals} and \ref{spectralf} 
are to be solved iteratively
and at each iteration the chemical potential $\mu$ is adjusted to fulfill
the condition on the density $\rho$
\begin{equation}
\label{densitycon}
\int\frac{d\omega}{2\pi}\int\frac{d^3p}{(2\pi)^3}A(p,\omega)f(\omega)=\rho \ .
\end{equation}

\begin{figure}

\centering
\includegraphics[width=0.5\textwidth]{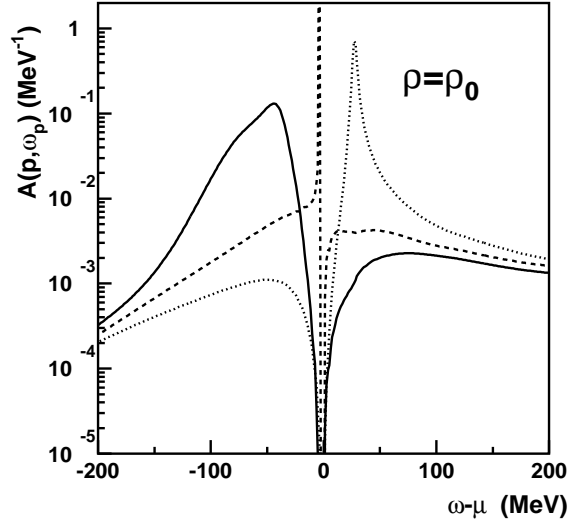}
\caption{The spectral function $A(p,\omega)$ as function of energy for 
$p=0,  255$ and $340$MeV (solid, dashed, and dotted lines respectively).}
\label{specfig}
\end{figure}

The spectral functions obtained in the self-consistent solution
consist of a quasiparticle peak and a broad background (Fig. \ref{specfig}).
As function of momentum the position of the peak in the spectral function 
follows approximately the quasiparticle dispersion relation
\begin{equation}
\omega_p=\frac{p^2}{ m} +{\rm Re}{\Sigma}(p,\omega_p) \ .
\end{equation}
The quasiparticle peak is very sharp for momenta close to the Fermi momentum.
It is a manifestation of the quasiparticle nature of  excitations close 
to the Fermi surface. Indeed we find that the single-particle width
$-2{\rm Im}{\Sigma}(p,\omega_p)$ is proportional to $(p-p_F)^2$ close to the 
Fermi momentum.
The background of the spectral functions extend far from the quasiparticle 
peak. The  part of the spectral function below the Fermi energy
leads  to nonzero occupancy for momenta above $p_F$ and
gives a large, negative contribution to the binding energy for all momenta.

The nucleon momentum distribution
\begin{equation}
n(p)=\int \frac{d\omega}{2 \pi} A(p,\omega)f(\omega)
\end{equation} is very different
different from the Fermi-Dirac distribution.
Momentum states below the Fermi momentum are depleted and a tail 
in the distribution $n(p)$ for large momenta appears.
The $T$-matrix approximation leads to 
 a Fermi liquid behavior in the normal phase,
with a jump in the fermion density of $Z_{p_F}=\bigg(1-\frac{\partial{\rm Re}
\Sigma(p_F,\omega)}{\partial \omega}|_{\omega=E_F}\bigg)^{-1}
\simeq .74$ at the Fermi momentum.
%The chemical and the corresponding Fermi momentum are fixed to satisfy 
%the condition \ref{denfix}.
In the calculation the Fermi momentum is fixed by the constraint 
(\ref{densitycon}) on the  total density.
For a conserving approximation the Fermi  momentum should be the same as
 the Fermi
momentum of a free fermion gas \cite{lw,baym}. 
This thermodynamical consistency relation is 
verified to a good accuracy by our calculations for the range of densities
studied.

\begin{figure}

\centering
\includegraphics[width=0.5\textwidth]{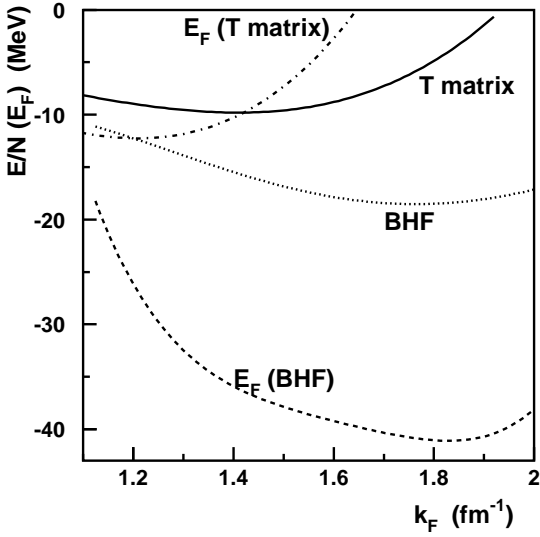}
\caption{The binding energy for the $T$-matrix (solid line)
and for the BHF (dotted line) calculations, and the Fermi energy
 for the $T$-matrix (dashed-dotted line)
and for the BHF (dashed line) calculations as 
function of the Fermi momentum.}
\label{befig}
\end{figure}

The energy per particle, in the case of only two-body interactions,
 can be obtained  from  the single-particle spectral function
\begin{equation}
\label{bin}
E/N=\frac{1}{\rho}
\int\frac{d^3p}{(2 \pi)^3}\frac{d\omega}{2\pi}
\frac{1}{2}\bigg(\frac{p^2}{2 m}+
\omega \bigg) A(p,\omega)f(\omega) \ .
\end{equation}
The binding energy per nucleon as function of the Fermi momentum
is presented 
in Fig. \ref{befig} for the self-consistent $T$-matrix approximation and
compared to results from $G$-matrix calculations using  the
continuous choice of the auxiliary potential.
The results of the $T$-matrix approach lie above to the  BHF 
binding energy for densities close to the phenomenological saturation point.
Since we know that further hole line corrections do not modify the 
continuous BHF results drastically, we get an assessment of the accuracy of the
$T$-matrix approach.
The higher the density the larger the
discrepancy becomes. 
Correspondingly the saturation point in the $T$-matrix approach is shifted to
lower densities ($\rho=1.4\rho_0$ instead of $2.4\rho_0$) 
and lower binding energies (the binding energy is reduced by $4{\rm MeV}$ at
$\rho_0$). 
Very similar results are found for
the equation of state of pure neutron matter \cite{jc2}. 
We note the the Hugenholz-Van Hove condition (\ref{hvheq})
 is very well satisfied.

\begin{figure}

\centering
\includegraphics[width=0.5\textwidth]{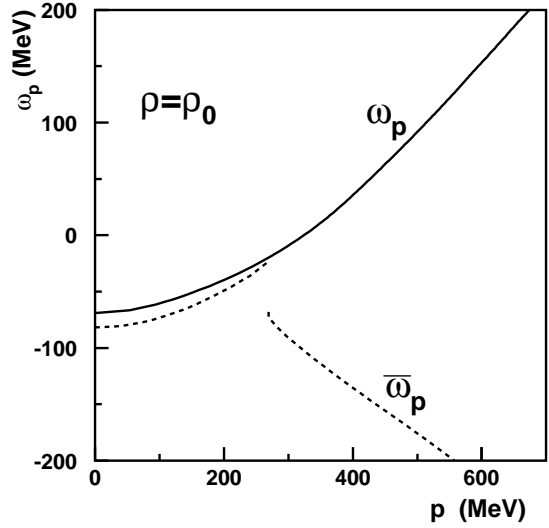}
\caption{The quasiparticle energy $\omega_p$ and the average
 energy $\overline{\omega}_p$  (\ref{aven}) as function of momentum.}
\label{omfig}
\end{figure}

The origin of the of the binding energy in the $T$-matrix 
approximation can be understood writing Eq. (\ref{bin}) as
\begin{equation}
E/N=\frac{1}{\rho}\int \frac{d^3p}{(2\pi)^3} n(p) \frac{1}{2}\left( 
\frac{p^2}{2m} + \overline{\omega_p}
\right) 
\end{equation}
with 
\begin{equation}
\label{aven}
 \overline{\omega_p}=\int \frac{d\omega}{2\pi} 
\omega A(p,\omega) f(\omega)/n(p) \ ,
\end{equation} 
whereas in  quasiparticle approaches it is
\begin{equation}
\label{qpbi}
E/N=\frac{1}{\rho}\int \frac{d^3p}{(2\pi)^3} n(p) \frac{1}{2}
\left( \frac{p^2}{2m} + \omega_p
\right) \ .
\end{equation}
In Fig. \ref{omfig} we compare the removal energy $\overline{\omega_p}$
to the quasiparticle energy $\omega_p$.
Due to a large contribution of the background strength of the spectral 
function lying below the quasiparticle peak the removal energy is 
much below the quasiparticle energy. On the other hand,
the single-particle energy $\omega_p$ in the $T$-matrix 
approximation is generally above the one obtained from $G$-matrix calculations.
It can be seen by comparing the Fermi energies in the two approximations 
in Fig. \ref{befig}. These differences explain why the
 Fermi energy in the
conserving $T$-matrix approximation is equal to the binding energy at 
the local saturation point, following the Hugenholz-Van Hove relation 
(\ref{hvheq}).
The binding energy is determined by $\overline{\omega_p}$ and
$E_F=\omega_{p_F}$ for the $T$-matrix calculation, whereas both quantities are 
determined by $\omega_p$ in the $G$-matrix scheme.

This paper presents the first results  self-consistent $T$-matrix 
calculation of saturation properties of nuclear matter 
with a realistic potential. The binding energy obtained is smaller than
 the BHF result with the continuous auxiliary potential.
We note that  a very similar shift in binding energy is observed in a BHF 
calculation, when including the rearrangement terms contribution to the
binding energy \cite{molinari}.
In Ref. \cite{molinari} by considering rearrangement term corrections to the
single particle energies and to the binding energy an improvement of the
fulfillment 
Hugenholz-Van Hove relation is found. The same can be observed in the
$T$-matrix approach, we destroy a bit the binding energies 
and improve considerably the single particle energies  
from the BHF approach to get the relation (\ref{hvheq}) right.
% and to the $G$-matrix
%calculation with three-hole lines contributions. 
%The off-shell properties of the spectral function are determined
%by the lowest partial waves (mainly $^3S_1-^3D_1$ and $^0S_1$) \cite{ja3}.
%Thus 
We expect that
 after inclusion of ring diagrams contributions, as well as  
higher partial waves and three body forces corrections, the results on the
saturation properties of nuclear matter of modern BHF approaches will be 
recovered. 
The calculation of these corrections is standard and not related 
to the $T$-matrix approach. On the other hand, the real advantage of the 
self-consistent $T$-matrix approximation shows itself in the single-particle
properties.  The quasiparticle energies lead to a Fermi energy
 consistent with the Hugenholz-Van Hove relation. The spectral function for 
large negative and positive energies can be calculated, 
with implications for the binding energy and applications for 
electron scattering. We find at normal nuclear density an effective mass
$m^{\star}=.9m$ and $Z_{p_F}=.74$. These numbers can be used in the 
estimation of reduced in medium cross sections and density of states at
the Fermi energy. Generally, $Z_p$ does not change much with density or with
 the details of the potential used, provided the lowest partial waves are
 described reasonably. We  confirm the thermodynamical consistency of
 the numerical solution of the $T$-matrix scheme \cite{jc}  for realistic
 interaction with several partial waves also.
%\cite{di1,ja3,jarostock,fermiliq}. 
 Finally let us note that the 
self-consistent $T$-matrix calculation can be straightforwardly
extended to finite temperatures.

%\bf Acknowledgments
\vskip .3cm
This work was partly supported by the KBN
under Grant No. 2P03B02019.

%\begin{references}

%\begin{figure}

%\centering
%\includegraphics[width=0.7\textwidth]{fermion.eps}
%\caption{Momentum distribution of nucleons for the  $T-$matrix calculation
%(solid line) compared to the free fermion distribution.}
%\label{fermionfig}
%\end{figure}

%\newpage
%\begin{figure}
%
%\centering
%\includegraphics[width=0.7\textwidth]{szer.eps}
%\caption{Imaginary part of the spectral function $-{\rm Im}\Sigma(p,\omega)$ 
%as function of energy for 
%$p=0, 1, 255$ and $420$MeV.}
%\label{szerfig}
%\end{figure}

\end{document}